\documentclass[a4paper,fleqn,usenatbib]{mnras}
\usepackage{newtxtext,newtxmath}
\usepackage[T1]{fontenc}
\usepackage{ae,aecompl}

\usepackage[flushleft]{threeparttable}
\usepackage{natbib}
\usepackage{graphicx}	
\usepackage{amsmath}	
\usepackage{amssymb}	
\usepackage{hyperref}
\usepackage{booktabs}
\usepackage{longtable}
\hypersetup{draft}

\title[Solar-like oscillations from CV spectra]{Coefficients of variation for detecting solar-like oscillations}

\author[K.~J.~Bell, S.~Hekker and J.~S.~Kuszlewicz]{
Keaton J.~Bell,$^{1,2}$\thanks{E-mail: bell@mps.mpg.de}
Saskia Hekker$^{1,2}$
and James S. Kuszlewicz$^{1,2}$
\\
$^{1}$Max-Planck-Institut f\"{u}r Sonnensystemforschung, Justus-von-Liebig-Weg 3, 37077 G\"{o}ttingen,
Germany \\
$^{2}$Department of Physics and Astronomy, Stellar Astrophysics Centre, Aarhus University,
Ny Munkegade 120, 8000 Aarhus C, Denmark
}

\date{Accepted XXX. Received YYY; in original form ZZZ}

\pubyear{2018}

\newcommand{\Kep}{\emph{Kepler}}

\newcommand{\numax}{$\nu_\textrm{max}$}

\begin{document}
\label{firstpage}
\pagerange{\pageref{firstpage}--\pageref{lastpage}}
\maketitle

\begin{abstract}
Detecting the presence and characteristic scale of a signal is a common problem in data analysis. We develop a fast statistical test of the null hypothesis that a Fourier-like power spectrum is consistent with noise. The null hypothesis is rejected where the local ``coefficient of variation'' (CV)---the ratio of the standard deviation to the mean---in a power spectrum deviates significantly from expectations for pure noise ($\mathrm{CV}\approx1.0$ for a $\chi^2$ 2-degrees-of-freedom distribution). This technique is of particular utility for detecting signals in power spectra with frequency-dependent noise backgrounds, as it is only sensitive to features that are sharp relative to the inspected frequency bin width. We develop a CV-based algorithm to quickly detect the presence of solar-like oscillations in photometric power spectra that are dominated by stellar granulation. This approach circumvents the need for background fitting to measure the frequency of maximum solar-like oscillation power, \numax. In this paper, we derive the basic method and demonstrate its ability to detect the pulsational power excesses from the well-studied APOKASC-2 sample of oscillating red giants observed by \Kep. We recover the cataloged \numax\ values with an average precision of 2.7\% for 99.4\% of the stars with 4 years of \Kep\ photometry. Our method produces false positives for $<1\%$ of dwarf stars with \numax\ well above the long-cadence Nyquist frequency. The algorithm also flags spectra that exhibit astrophysically interesting signals in addition to single, solar-like oscillation power excesses, which we catalog as part of our characterization of the \Kep\ light curves of APOKASC-2 targets. 
\end{abstract}




\section{Introduction}

Our ability to constrain the properties of stars has been recently revolutionized by space-based missions dedicated to obtaining extended, precise time series photometry such as CoRoT \citep{Baglin2006} and \Kep\ \citep{Borucki2010}. The conditions of stellar interiors have been especially illuminated by detections of solar-like oscillations in thousands of red giant stars with convective outer envelopes \citep[e.g.,][and references therein]{Hekker2009,Bedding2011,Mosser2011b,Mathur2011a,Stello2013,Yu2018}. Turbulent motions in these layers stochastically excite globally damped stellar eigenmodes with eigenfrequencies near the convective turnover timescale \citep{Goldreich1977}.  These pulsations propagate through the stellar interior to manifest as incoherent photometric variations, which can be revealed through Fourier analysis of light curves. Detecting the presence of oscillations in a light curve gives the basic indication that a target is suited for asteroseismology \citep[e.g.,][]{Chaplin2014,Miglio2016}.  The characteristic frequency of maximum power of these oscillations, \numax, is a precise tracer of the surface gravity in red giants \citep[e.g.,][]{Brown1991,Kjeldsen1995,Hekker2013}. Asteroseismic determinations of this fundamental stellar parameter propagate to tight constraints on stellar and exoplanetary masses and radii \citep[e.g.,][]{Ballard2014,Campante2015}. \citet{Hekker2017} give a recent review of the field of giant star seismology and the science that is enabled by detecting solar-like oscillations.

The same turbulent convection that drives solar-like oscillations in red giants produces additional signatures from granulation that dominate the power spectra \citep{Harvey1985}. These features can be described by multiple ``super-Lorentzians'' that decay to high frequency as $\nu^{-4}$ \citep{Aigrain2004,Kallinger2014}. The oscillation signals are comparatively sharp---the widths of their Lorentzian profiles \citep[$\sim0.1\,\mu$Hz;][]{Baudin2011} are inversely proportional to the mode lifetimes \citep[e.g.,][]{JCD1989}. To peakbagging efforts that aim to precisely measure pulsation mode characteristics, the granulation effectively contributes frequency-dependent noise to the power spectrum backgrounds. Obtaining and removing a global fit to these backgrounds is often the most computationally expensive step in asteroseismic analyses. However, this process could be greatly accelerated by independent determinations of \numax, as these are tightly correlated with the background component parameters through empirical scaling relations \citep{Kallinger2014}.

Despite the challenges introduced by the background, spotting the power excesses of solar-like oscillations is not considered a difficult problem for the trained eye.  We develop a new method for detecting solar-like oscillations that mimics the approach of an expert, who would prefer to visually inspect the logarithm of the power spectrum. In this representation, the power is distributed about a background that varies over orders of magnitude as a band with near-constant width. Near \numax, the oscillations broaden the logarithmic power spectrum locally. We search for local excesses in power spectrum scatter by evaluating the ``coefficient of variation'' (CV; standard deviation divided by the mean) in limited frequency ranges. We reject the null hypothesis that a bin contains only noise where this CV value is sufficiently large. This method is successful in detecting sharp spectral features, while being relatively insensitive to slowly varying (in frequency) noise backgrounds. For solar-like oscillators, this enables us to quickly detect and measure the \numax\ of pulsational power excesses without having to fit the granulation background.

Determining which targets of large time domain surveys exhibit oscillations is the crucial first step toward widespread asteroseismic characterization of stellar populations. While they have already yielded a multitude of important asteroseismic results, the CoRoT, \Kep\, and K2 \citep{Howell2014} data sets remain treasure troves of unidentified solar-like oscillators. Over the next two years, the \emph{Transiting Exoplanet Survey Satellite} \citep[\emph{TESS};][]{Ricker2015} promises to measure photometry for upwards of 20 million stars. By quickly evaluating which targets do or do not exhibit solar-like oscillations, the CV method stands to improve out efficiency in processing these data.

We develop the statistics for the CV spectrum in general terms in Section~\ref{sec:metric}, including a significance criterion.  In Section~\ref{sec:rgs}, we use the empirical seismic scaling relations to tune a CV-based algorithm for quickly detecting and measuring \numax\ of solar-like oscillation power excesses. We demonstrate that our method yields the expected results for most dwarf stars (null detections; Section~\ref{sec:dwarfs}) and giants (power excess detections at accurate \numax\ values; Section~\ref{sec:apokasc}). Our method identifies stars that exhibit other interesting features in addition to solar-like oscillation power excesses, and these objects are flagged in our characterization of the APOKASC giants \citep{Pinsonneault2014,Pinsonneault2018} in Appendix~\ref{app:tab}. In Appendix~\ref{app:multi}, we confirm from visual inspection 30 light curves  that appear consistent with blends of multiple solar-like oscillators.

\section{The Coefficient of Variation (CV) Metric}
\label{sec:metric}

Fourier power spectra are commonly employed for detecting frequencies of intrinsic variability in time series data. Power from deterministic or stochastic signals is distributed differently than power that represents background noise.  We aim to exploit these differences to quickly distinguish the signals without having to precisely model the background.  We develop an approach based on the approximate statistical properties of the background that we find to be a useful tool for identifying signals in the presence of frequency-dependent background noise. We will demonstrate this by detecting solar-like oscillations from time series photometry in later sections.  In this section, we derive the coefficient of variation method in general terms, as it is widely applicable beyond the domain for which it has been crafted here.

In the Fourier transform of a time series of regularly sampled Gaussian white (independent and uncorrelated) noise of duration $T$, the real and imaginary components sampled at independent frequencies separated by $1/T$ are both Gaussian distributed.  For time series that contain a large number of samples, the central limit theorem supports this result for arbitrary noise distributions. The power spectrum formed by summing the squares of these terms therefore consists of independent and identically distributed (i.i.d.) random variates from a $\chi^2$ probability density function (p.d.f.) with 2 degrees of freedom\footnote{
	\label{chi2}Because we must estimate the variance from the time series, the power \emph{measured} in a pure-noise spectrum is not exactly $\chi^2_2$ distributed; however, this difference is negligible in the limit of a large number of observations \citep{SC1998,Frescura2008}.}
\citep[d.o.f.; e.g.,][Section 9.4.2]{Robinson2017}. This $\chi^2_2$ distribution equals the exponential distribution with a rate parameter of 1/2:
\begin{equation}
F(x) = 
\begin{cases} 
\frac{1}{2}e^{-x/2} & x\geq 0 \\
0 & x < 0\,.
\end{cases}
\end{equation}

The power spectrum computed from the Lomb-Scargle periodogram is formulated to produce the same $\chi^2_2$ noise statistics for unevenly sampled data\footnote{
	\label{foot:indep}Structure in the spectral window for unevenly sampled data causes there to exist no frequencies at which the periodogram powers are truly independent \citep[see, e.g.,][]{Frescura2008}. For our purposes, we proceed to treat power sampled at $1/T$ as effectively independent for nearly evenly-spaced data.}
\citep{Scargle1982}. We aim to exploit the behavior of this distribution to test the null hypothesis that a region of a power spectrum contains only noise---namely, that the underlying mean and standard deviation are equal\footnote{Generally, the mean and standard deviation of a $\chi^2$ distribution with $f$ d.o.f.\ are $f$ and $\sqrt{2f}$, respectively.}. The ratio of the standard deviation to the mean---called the ``coefficient of variation'' (CV)---for a $\chi^2_2$ distribution is therefore equal to one.  We expect to measure a CV close to this value in the power spectrum of pure white noise.  The presence of any additional, localized signal will have the effect of increasing the CV, often to some statistical significance.

\begin{figure}
	\includegraphics[width=1\columnwidth]{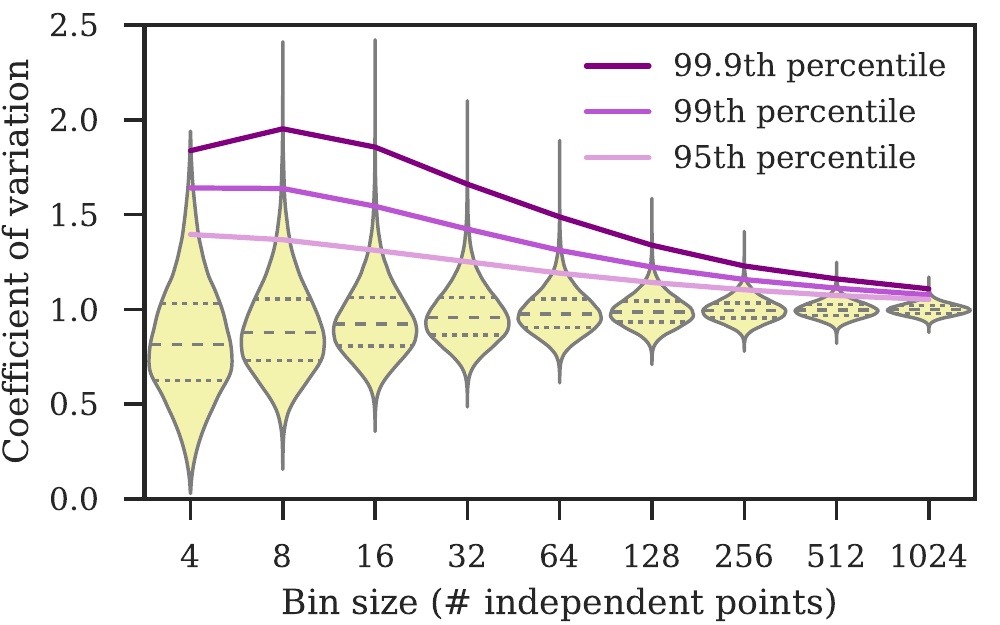}
	\caption{Violin plot showing kernel density estimations of the CV measurements obtained from different sample sizes of simulated $\chi^2_2$ i.i.d.s. Solid lines connect the 95th, 99th, and 99.9th percentiles of each, corresponding to 5\%, 1\%, and 0.1\% false alarm probabilities for pure noise in the power spectra, respectively. The dashed and dotted lines within each distribution indicate the medians and interquartile ranges.}
	\label{fig:FAP}
\end{figure}

\begin{table}
	\centering
	\caption{CV false alarm probabilities.}
	\label{tab:FAPs}
	\begin{tabular}{rccc}
		\hline
		bin size & 5\% FAP & 1\% FAP & 0.1\% FAP\\
		(\# pts) & (CV) & (CV) & (CV)\\
		\hline
		4    & 1.397 & 1.640 & 1.830 \\
		8    & 1.368 & 1.632 & 1.942 \\
		16   & 1.314 & 1.540 & 1.865 \\
		32   & 1.253 & 1.421 & 1.664 \\
		64   & 1.191 & 1.311 & 1.486 \\
		128  & 1.141 & 1.225 & 1.341 \\
		256  & 1.102 & 1.157 & 1.227 \\
		512  & 1.073 & 1.110 & 1.157 \\
		1024 & 1.052 & 1.077 & 1.108 \\
		\hline
	\end{tabular}
\end{table}

In reality, correlated stochastic noise is more common than white noise; however, these backgrounds often show only gradual frequency dependences, with the noise still distributed as $\chi^2_2$ about a non-flat limit spectrum \citep[as for solar-like oscillators;][]{Duvall1986,Anderson1990}.  In the limit of increasingly narrow frequency ranges, the spectrum will appear \emph{locally} white. Comparing the CV measured in a frequency range to expectations for $\chi^2_2$ noise can reveal spectral features that are sharp relative to the inspected bin width, while being insensitive to background features that vary over longer frequency scales.  Because the CV calculation normalizes the scatter in the power spectrum by the local mean, it does not require a precise characterization of the overall noise background to identify sharp features of interest.

In practice, we rely on direct simulations to derive a significance criterion for rejecting the null hypothesis that a measured CV arose purely from $\chi^2_2$ noise.  While the $\chi^2_2$ distribution does have a characteristic CV of unity, the measured standard deviation divided by the measured mean is a biased estimator of the CV for two reasons: (1) the standard deviation is biased (as a nonlinear transformation of an unbiased estimator: the variance); and (2) the variance and mean of i.i.d.s are not independent for non-Gaussian underlying p.d.f.s \citep{Geary1936}. Closed forms for the p.d.f.s of variance measured for generic sample sizes drawn from exponential distributions are not known \citep[e.g.,][]{Lam1980}. 

The distributions of measured CVs simulated for nine different sample sizes containing between 4 and 1024 i.i.d.\ random $\chi^2_2$ variates are displayed in Figure~\ref{fig:FAP}. Each distribution is determined from 100,000 sets of randomly generated values. The distributions converge with increasing sample size toward Gaussians centered on one, as expected from the central limit theorem.  The solid lines mark the 95th, 99th, and 99.9 percentiles of each distribution. These correspond to 5\%, 1\%, and 0.1\% false alarm probabilities (FAPs) that a detection above these thresholds was caused purely by $\chi^2_2$ noise in the data. FAPs for bins of generic size can be interpolated from the values provided in Table~\ref{tab:FAPs}. If multiple independent frequency bins are considered, the probability of a CV value in any bin exceeding these thresholds due solely to noise increases with the number of bins examined.

\section{CV Spectra For Detecting Solar-like Oscillations in Giants}
\label{sec:rgs}

Having presented a statistical metric for detecting sharp spectral features in the presence of frequency-dependent background noise, we now outline a CV-based strategy for detecting solar-like oscillations in red giants and measuring \numax. 
The following method is specifically designed for long-cadence \Kep\ light curves. It is validated by null results for dwarf stars with super-Nyquist $\nu_\mathrm{max} > 500\,\mu$Hz, and by our recovery of cataloged \numax\ values for the APOKASC giants \citep{Pinsonneault2014,Pinsonneault2018}. This is just one of many possible CV-based algorithms, and it could easily be adjusted for other data sets or to be more/less susceptible to false positives/negatives.

The pulsation signatures in the power spectra of solar-like oscillators fall under an approximately Gaussian envelope centered on the frequency of maximum oscillation power, \numax. Various other observational characteristics of red giant power spectra are tightly correlated with \numax, as described by a set of empirical power laws \citep[e.g.,][]{Mosser2010,Mosser2012,Mathur2011,Kallinger2014}. These scaling relations set our expectations about the signatures of solar-like oscillations and inform our strategy for computing CV spectra that will best reveal their presence.

\begin{figure*}
	\includegraphics[width=1.78\columnwidth]{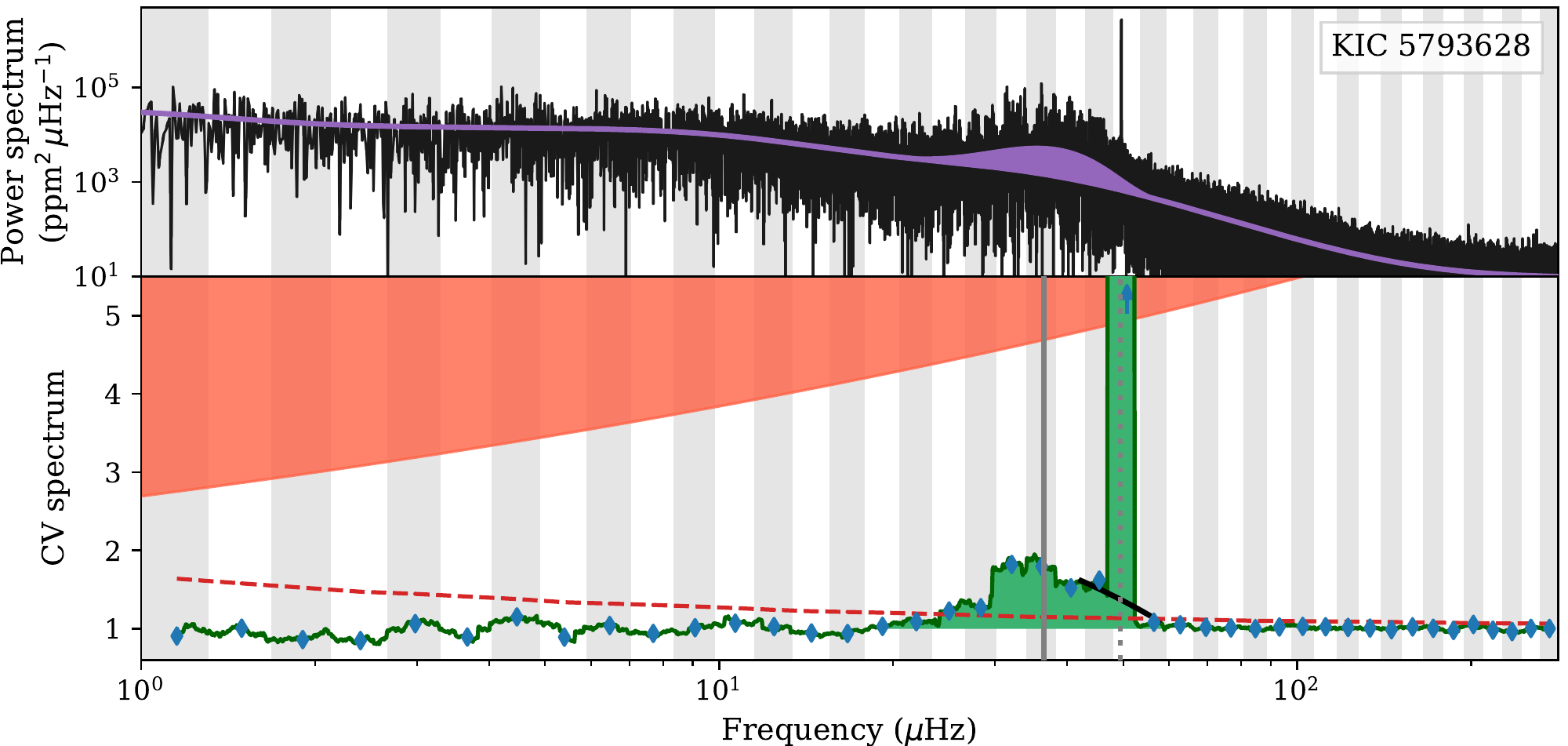}
	\caption{{\sc Top:} Power density spectrum of the APOKASC giant KIC\,5793628 on a log-log scale. A spike exceeding 2.7$\times$$10^6$\,ppm$^2\mu$Hz$^{-1}$ falls near the pulsation signals. A background model fit to the power spectrum following Kuszlewicz et al.\ (in prep.; fit with spike excluded) consisting of three super-Lorentzians, white noise, and a Gaussian power excess (filled) is overplotted. Gray and white bands indicate the 43 independent bins that scale with $\Delta\nu$ in both panels (see text). {\sc Bottom:} Corresponding CV spectrum.  The dashed red curve is the 0.1\% FAP significance threshold.  The blue diamonds mark the CV measurements from independent bins, with the arrow indicating one value above the plotted range. The oversampled CV spectrum is dark green. The spike causes the oversampled CV spectrum to exceed an upper expectation limit for solar like oscillations (shaded red), and the black line segment displays the interpolation over this feature. The thick gray line indicates the \numax\ value that we measure from the CV spectrum, and the dotted gray line is the value from APOKASC-2 (\citealt{Pinsonneault2018}; coincident with the large spike in this case; see Section~\ref{sec:apokasc}).  The oversampled spectrum is filled where it is continuously above a level of $\mathrm{CV}=1$ surrounding the reported detection of solar-like oscillations.}\label{fig:CVexample}
\end{figure*}

The calculation of CV spectra requires binning of the power spectra, and we want to adopt a binning scheme that well-captures the oscillatory signal near any \numax. The pressure-modes of solar-like oscillations are distributed according to a ``universal pattern'' \citep{Mosser2011}, with subsequent radial overtones of the same spherical degree exhibiting a nearly even frequency spacing \citep{Tassoul1980}.  Many studies have empirically constrained the asteroseismic scaling relation for this ``large frequency separation'', $\Delta\nu$,  all finding roughly $\Delta\nu\propto\nu_\mathrm{max}^{0.75\textup{--}0.78}$ \citep[e.g.,][]{Hekker2009,Stello2009,Huber2010,Mosser2012,Yu2018}. Since we expect each acoustic radial order to exhibit pulsation signals within the power excess, we compute CV values in bins with widths that scale with the expected $\Delta\nu$ at each searched frequency. The power-law form of the scaling relation makes it convenient to consider our spectra with log-frequency scales. The ``large logarithmic frequency separation'', $\Delta\log{\nu}$ \citep[in units of dex;][]{Cox1951}, decreases linearly with $\log{\nu_\mathrm{max}}$ from $\Delta\log{\nu}\approx0.067$\,dex at $\nu_\mathrm{max}=10\,\mu$Hz to $\Delta\log{\nu}\approx0.033$\,dex at $\nu_\mathrm{max}=200\,\mu$Hz.

The FAP significance thresholds given in Section~\ref{sec:metric} are only independently applicable to non-overlapping bins.  We numerically calculate edges of 43 contiguous bins that span a range between $1\,\mu$Hz and the Nyquist frequency of the long-cadence \Kep\ data, $283.2\,\mu$Hz \citep[following the most recent relation from][$\Delta\nu\approx 0.267\times\nu_\mathrm{max}^{0.764}$]{Yu2018}. These bin edges are indicated in Figure~\ref{fig:CVexample}, which displays a representative CV spectrum in comparison with the original power spectrum for the red giant KIC\,5793628. The width of each bin is 1.029 times the average expected $\Delta\log\nu$. We compare the CV values calculated in these bins to the 0.1\% FAP threshold corresponding to the number of independent power spectrum frequencies they contain. As indicated by the dashed line in the bottom panel of Figure~\ref{fig:CVexample}, lower-frequency bins generally have higher significance thresholds because their narrower linear frequency ranges contain fewer independent frequencies of the power spectrum.

While the CV spectrum calculated with non-overlapping bins is convenient for significance testing, it sparsely samples pulsational power excesses.  Therefore, we compute a second CV spectrum with 2{\,}000 overlapping bins, evenly spaced in log-frequency over the same range and following the same scaling relation for $\Delta\nu$. We refer to this as the ``oversampled'' CV spectrum, in contrast to the former ``independent'' spectrum. The oversampled spectrum is displayed in green in the bottom panel of Figure~\ref{fig:CVexample}. This resembles a smoothed, background-removed power spectrum that enables precise determinations of \numax.

Before further analysis, we search our oversampled CV spectra for contaminating signals that may impact our measurements.  Signatures of, e.g., binarity, rotation, instrumental noise, or blends with classical pulsators, often introduce large spikes to our CV spectra. By inspection of the maximum CV values near the cataloged \numax\ of APOKASC stars (Section~\ref{sec:apokasc}), we determine a rough upper limit for CV values expected from solar-like oscillations: CV$_\mathrm{max}\approx 2.69\times (\nu/\mu\mathrm{Hz})^{0.154}$. We linearly interpolate across any regions of the oversampled spectrum that share a dependence with CV values that exceed this threshold. The black line segment in the bottom panel of Figure~\ref{fig:CVexample} shows the interpolation across such a spike. Cleaning the spectrum in this way enables robust \numax\ measurements, even when a large contaminating signal falls near the power excess. Our measurements do not lose precision in the few cases where CVs from solar-like oscillations themselves exceed this threshold.

While we expect solar-like oscillations to span multiple acoustic radial orders within the pulsational power excesses, the lower-amplitude signals near the edges of the Gaussian power envelope may not be sufficient to increase the CV above the FAP threshold. Nevertheless, they are likely to exceed the limiting expectation value of 1.0.  For $\nu_\mathrm{max} < 100\,\mu$Hz, \citet{Mosser2012} find that the full-widths at half-maximum (FWHM) of the oscillation envelopes follow $\delta\nu_\mathrm{env}\approx 0.66\nu_\mathrm{max}^{0.88}$. This trend is observed to be shallower above 100\,$\mu$Hz. We define our algorithm to expect $\delta\nu_\mathrm{env}$ to scale with $\Delta\nu$, fixing $\delta\nu_\mathrm{env} \approx 4.2\Delta\nu$ for $\nu_\mathrm{max} > 100\,\mu$Hz. We adopt a candidate solar-like oscillation power excess if its independent CV spectrum contains values between the 0.1\% FAP threshold and the empirical CV$_\mathrm{max}$ limit in a region where the oversampled CV spectrum remains sustained above 1.0 for the span of at least $\delta\nu_\mathrm{env}$.

After accepting that a frequency range likely contains \numax, we refine the measurement by computing local moments of the oversampled CV spectrum. Searching over the full \mbox{CV $>1$} power excess, we identify the $\delta\nu_\mathrm{env}$-wide frequency window containing the largest average CV (effectively a second smoothing step). We then record the \numax\ value to be the center-of-mass of the oversampled CV spectrum (first moment divided by the integrated CV) in that window.

Our algorithm can naturally return multiple \numax\ values when the time series captures blended sources.  We require that these are sufficiently resolved (separated by at least $\delta\nu_\mathrm{env}$) to all be reported. We also record whether the independent CV spectra contain additional significant peaks outside the expected Gaussian $3\sigma$ ranges surrounding accepted \numax\ values, which may be scientifically interesting.

Signals from intrinsic oscillations with frequencies that exceed the Nyquist (283.2\,$\mu$Hz for \Kep\ long cadence) will be reflected back into the sub-Nyquist regime \citep[e.g.,][]{Chaplin2014b}.  
This folding of the power spectrum introduces systematics in our measurements for $\nu_\mathrm{max}\gtrsim245.6\,\mu$Hz. The span of \mbox{CV $>1$} required at this limit is also restricted, so false positives are somewhat more likely. For stars with \numax\ above the Nyquist, our algorithm is only sensitive to the sub-Nyquist aliases.

\begin{table}
	\centering
	\caption{Bit field flag definitions.}
	\label{tab:flags}
	\begin{tabular}{lll}
		\hline
		bit & value &  meaning\\
		\hline
		1 & 1 & additional significant, independent CVs \\
		2 & 2 & CV spike exceeds solar-like expectation \\
		3 & 4 & more than one \numax\ candidate detected \\
		4 & 8 & FliPer \numax\ disagreement \\
		5 & 16 & \numax\ candidate near Nyquist frequency \\
		\hline
	\end{tabular}
\end{table}

As a final consistency check, we compare our \numax\ values to independent estimates based on the stellar granulation.  Because the granulation signatures that dominate the power in the light curves are correlated with surface gravity \citep[e.g.,][]{Mathur2011}, the variance can be used as a rough proxy for \numax\ \citep{Hekker2012}. The most recent tool for validating \numax\ in this way is the ``FliPer'' metric (\citealt{Bugnet2017}\footnote{\citet{Bugnet2018} trained a random forest regressor to estimate surface gravity or \numax\ based on a combination of FliPer values computed over multiple frequency ranges with effective temperature and \Kep\ magnitude. We use the scaling relation formalism from \citet{Bugnet2017} in this work.}). This empirical relationship scales from the mean of the power density spectrum, corrected for the photon noise.  To be less affected by near-Nyquist signal, we use \citeauthor{Bugnet2017}'s recommended prescription based on the parameterized instrumental noise level from \citet{Jenkins2010}. We flag \numax\ results that disagree with the  FliPer estimate by more than 0.5\,dex. Potential causes of such disagreements include: false positive detections of \numax; detection of aliases of super-Nyquist power excesses; or, the \numax\ and variance not agreeing with scaling relations, e.g., for blended sources. We note that the $\sim$1-dex-wide acceptable \numax\ ranges estimated from FliPer \citep[][Figure 3]{Bugnet2017} span about 40\% of the log-frequency range that our CV algorithm searches.

In the end, our CV analysis of a power spectrum outputs any (possibly multiple) \numax\ measurements. In addition, we flag a number of conditions to indicate that the measurements may be suspect or that the spectrum may have additional features of scientific interest. Each bit in the binary representation of a single returned value represents the condition of each of five flags, as defined in Table~\ref{tab:flags}.

\section{False positive test on \Kep\ dwarf sample}
\label{sec:dwarfs}

We have designed an algorithm to detect solar-like oscillation signatures under the hypothesis that the frequency-dependent granulation backgrounds do not mimic pulsational power excesses in CV spectra. We test this by applying it to long-cadence light curves of 391 dwarfs and sub-giants with $\nu_\mathrm{max} > 500\,\mu$Hz measured from short-cadence \Kep\ data by \citet[][]{Chaplin2014}. 
The pulsations of these stars are well above the Nyquist frequency, and their sub-Nyquist aliases will be suppressed below observational limits by the 30-minute \Kep\ exposures. The power spectra of these data still have coloured granulation backgrounds and potentially other contaminating signals.

We utilize the preprocessed light curves available on the \Kep\ Asteroseismic Science Operations Center (KASOC) website\footnote{\label{kasoc}\url{http://kasoc.phys.au.dk/}}. 
The \Kep\ light curves used in this work have been corrected following \citet{Handberg2014}.

All power spectra in this work were calculated from unweighted time series with the fast implementation of the Lomb-Scargle periodogram from the {\sc gatspy} package \citep{Vanderplas2016}. We rescale the spectra by the variance of the light curve to obtain units of ppm$^2\mu$Hz$^{-1}$ (ppm is parts per million) that agree with Parseval's theorem. These units are useful for computing \numax\ estimates from FliPer for validation, yet constant multiplicative offsets to the power spectra do not affect the CV spectra.

The KASOC light curves omit data that were acquired during \Kep\ reaction wheel desaturation events that occurred every three days.  The regularity of these missing points introduces a significant comb of aliases into the spectral window, causing power to be broadly redistributed through frequency space.  The resulting backgrounds in the power spectra of such time series generally do not follow $\chi^2_2$ distributions, and the correlated features every 3.86\,$\mu$Hz can increase the CV above our significance threshold.  The effects of the window function on stellar oscillation spectra were analyzed by, e.g., \citet{Garcia2014}. Linear interpolation across single missing observations improves the spectral window sufficiently to avoid most spurious CV detections\footnote{No gap interpolation was performed on the pre-computed APOKASC power spectra available on the KASOC website$^\mathrm{\ref{kasoc}}$, so they are not suited for this analysis.}.

\begin{figure}
	\centering
	\includegraphics[width=1\columnwidth,trim={0 2mm 0 0},clip]{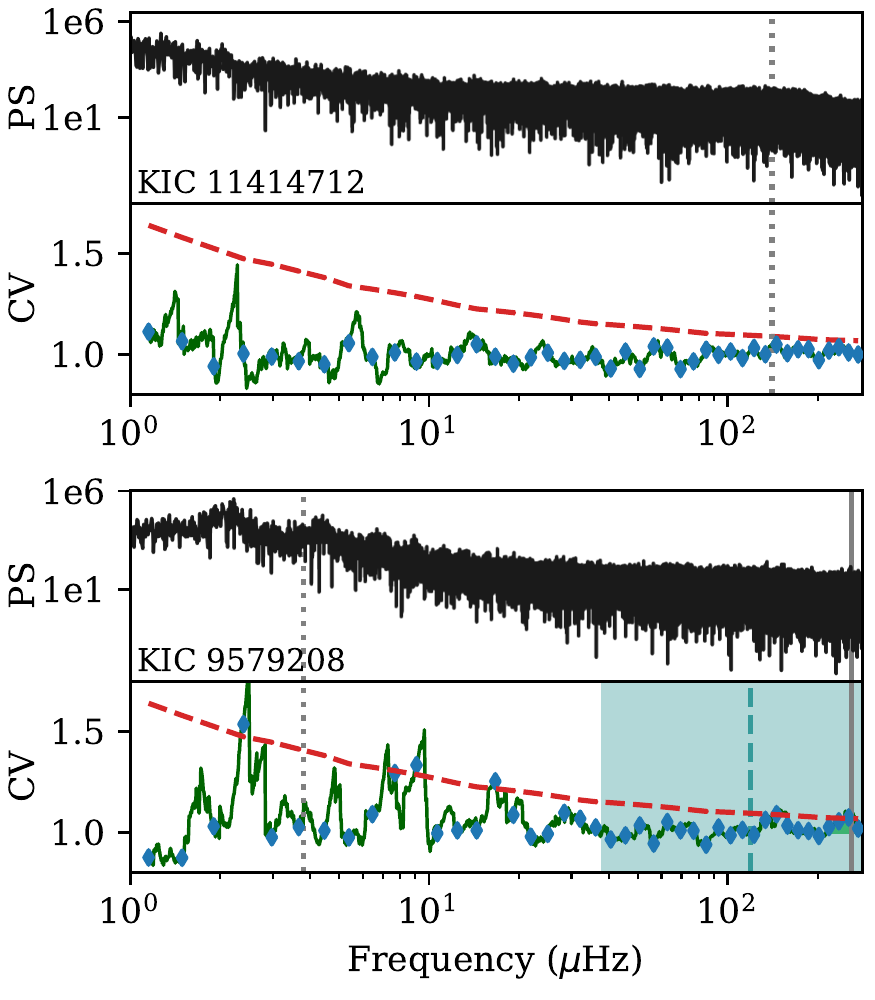}
	\caption{Power and CV spectra of two dwarf stars: KIC\,11414712 (top) and KIC\,9579208 (bottom). The power spectra are in units of ppm$^2\,\mu$Hz$^{-1}$. The vertical dotted lines indicate the sub-Nyquist aliases of their intrinsic values of $\nu_\mathrm{max} \approx 707\,\mu$Hz  and $\nu_\mathrm{max} \approx 1129\,\mu$Hz, respectively \citep{Chaplin2014}. The CV spectra from independent bins are plotted as blue diamonds, and the oversampled CV spectra are displayed in dark green (see text).  The red dashed curves give the 0.1\% FAP significance thresholds. 
		The power spectrum of KIC\,11414712 exhibits a frequency-dependent background, but no significant deviations from expectations for noise are present in the CV spectrum.
		KIC\,9579208 is the only dwarf target that yields a false positive detection (filled green) that agrees with the acceptable range of \numax\ values given the FliPer metric (shaded blue).}
	\label{fig:fp}
\end{figure}

No features in the CV spectra of 369 out of 391 dwarfs examined match the height and width requirements to be considered candidate solar-like oscillation power excesses. The top panels of Figure~\ref{fig:fp} display the power and CV spectra of a representative dwarf star for which our algorithm does not detect any pulsations: KIC\,11414712. Of the 22 candidate \numax\ values returned, the majority are at frequencies below 10\,$\mu$Hz and correspond to localized rotation signatures. However, the FliPer \citep{Bugnet2017} validation test rejects all but one of these candidates. We display the power and CV spectra of this target, KIC\,9579208, in the bottom panel of Figure~\ref{fig:fp}. While FliPer is not a very precise proxy for \numax, it is very useful for vetting candidate power excess detections. The \numax\ returned for this target is also flagged for falling in the ``near-Nyquist'' ($\nu_\mathrm{max}\gtrsim245.6\,\mu$Hz) range where a relaxed width requirement makes false positives more likely. We conclude that fewer than 1\% of \numax\ detections returned by our algorithm that agree with the estimate from FliPer are false positives.

\section{Application to the APOKASC Giants}
\label{sec:apokasc}

The second APOKASC catalog \citep[APOKASC-2;][]{Pinsonneault2018} characterizes 6676 evolved stars with both \Kep\ light curves and APOGEE spectra \citep{Pinsonneault2014}. It includes \numax\ values combined from an ensemble of pipelines.  We use these as a benchmark for evaluating the sensitivity, accuracy, and precision of our method of detecting \numax\ from CV spectra.

We apply the CV method to the 6656 APOKASC-2 stars with preprocessed light curves available from KASOC.
APOKASC-2 provides seismic measurements (\numax) for 6556 of these. Where available on KASOC, we inspect the short-cadence (1-minute) light curves to identify stars with intrinsic \numax\ values above the long-cadence Nyquist frequency of 283.2\,$\mu$Hz \citep[e.g.,][]{Chaplin2014b}. To test our method in this regime, we reflect the cataloged sub-Nyquist \numax\ aliases across the Nyquist frequency to recover the intrinsic \numax\ values of five super-Nyquist pulsators: KIC\,10394814, KIC\,6430804, KIC\,4351319, KIC\,7341231, and KIC\,3329196.

\begin{figure}
	\centering
	\includegraphics[width=1\columnwidth,trim={0 0 0 0},clip]{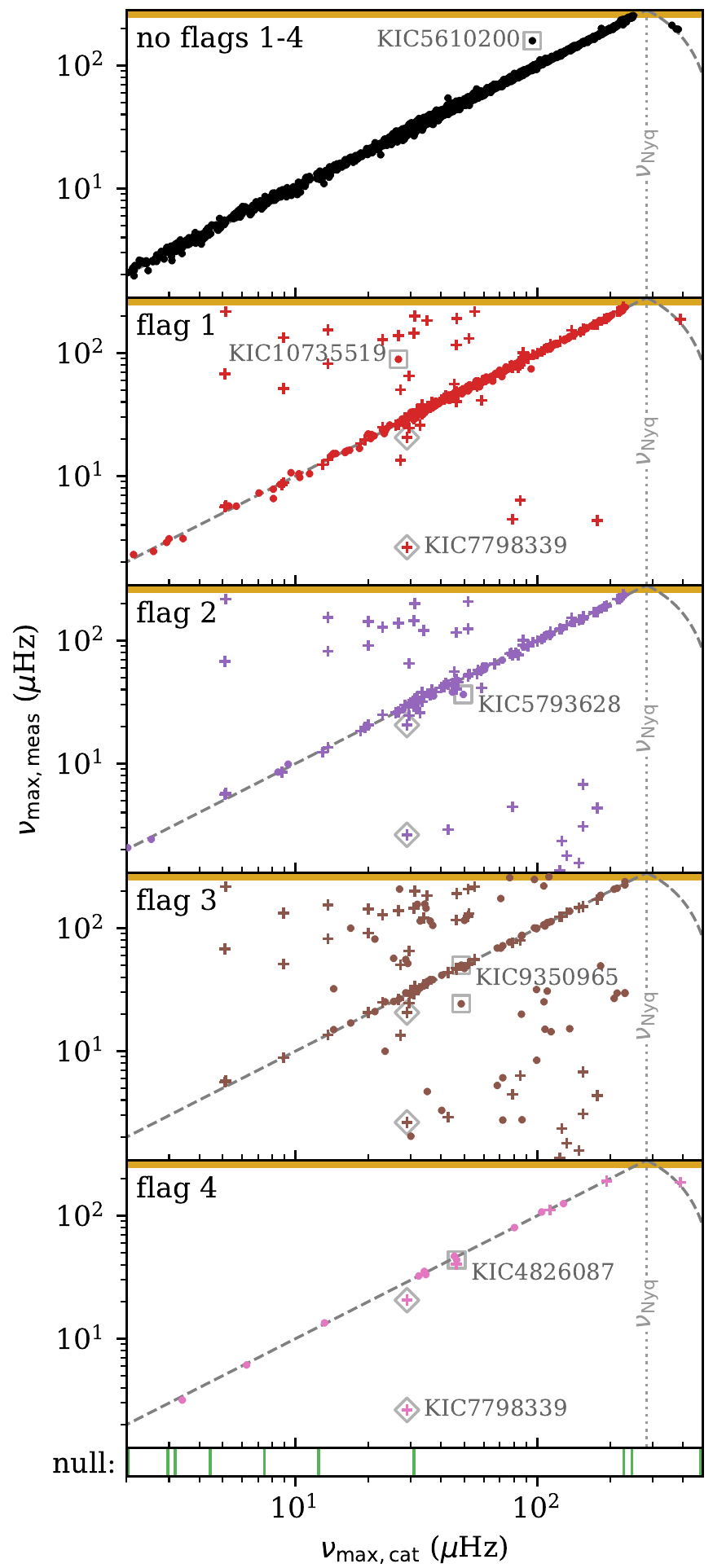}
	\caption{Comparison between the \numax\ values measured by our algorithm and the values from APOKASC-2 for the 4080 longest KASOC-processed \Kep\ light curves. The results are separated by the flags returned by our algorithm, which are defined in Table~\ref{tab:flags}. Stars marked with crosses are present in multiple panels. Stars with multiple \numax\ values returned are represented for each detection (flag 3).  Flag 5 is set for stars with \numax\ measured in the orange region near the Nyquist limit. The dashed line marks exact agreement, accounting for reflection above the Nyquist frequency, which is indicated with a dotted line.  Vertical bars in the bottom panel mark the cataloged \numax\ values of stars for which our algorithm does not return a \numax\ detection. Objects with spectra displayed as examples in Figures~\ref{fig:CVexample} and \ref{fig:examples} are outlined with boxes and labeled (see text for discussion).} 
	\label{fig:results}
\end{figure}

Our algorithm returns \numax\ measurements for 99.7\% of the 4080 longest light curves that APOKASC-2 reports \numax\ for---those spanning $>3.5$\,years with $>85\%$ duty cycle. We compare our detections to the APOKASC-2 \numax\ values for these stars in Figure~\ref{fig:results}. The results are separated into panels based on the flags returned (defined in Table~\ref{tab:flags}). Stars can appear in multiple panels if more than one flag is set. Stars with flag 3 set appear multiple times per panel (marked with crosses), once per candidate \numax\ value. Detections in the near-Nyquist region are also flagged (flag 5), and we consider these separately from the breakdown by flags 1--4.  Seismic measurements for most near-Nyquist stars were omitted in APOKASC-2.

\begin{figure}
	\centering
	\includegraphics[width=.9\columnwidth]{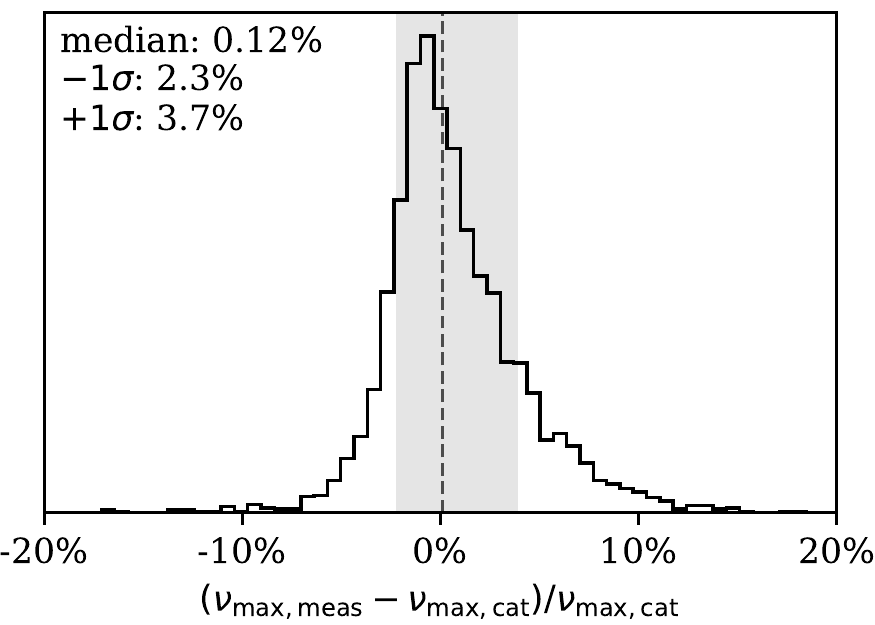}
	\caption{Histogram of the percentage difference between the APOKASC-2 cataloged \numax\ values and the results from the CV method for sub-Nyquist stars that do not return any flags.} 
	\label{fig:precision}
\end{figure}

The 3634 (89.1\%) \numax\ measurements that are not accompanied by flags 1--4 are in close agreement with the cataloged values, accounting for reflection at the Nyquist frequency (top panel of Figure~\ref{fig:results}). Figure~\ref{fig:precision} displays the histogram of the percentage difference between measured and cataloged \numax\ for sub-Nyquist stars that do not return any flags. The median difference is 0.12\%, and 68.3\% of our results agree within 3.0\% ($\sigma$) of the \numax\ values from APOKASC. The uncertainties quoted by APOKASC-2 imply that their \numax\ measurements are distributed about the intrinsic values with $\sigma = 1.3\%$. Subtracting this in quadrature from the $\sigma$ of the differences between measured and cataloged \numax\ values for these stars, we conclude that our \numax\ measurements have an average precision of 2.7\%. This includes both our intrinsic measurement errors and any systematic differences, which are typically of-order 1\% for \numax\ values output by different pipelines \citep[e.g.,][]{Pinsonneault2018}. We detect a \numax\ within 20\% of the cataloged value (i.e., the same power excess) for 99.4\% of stars, regardless of flags.

\begin{figure}
	\centering
	\includegraphics[width=1\columnwidth,trim={0 0 0 0},clip]{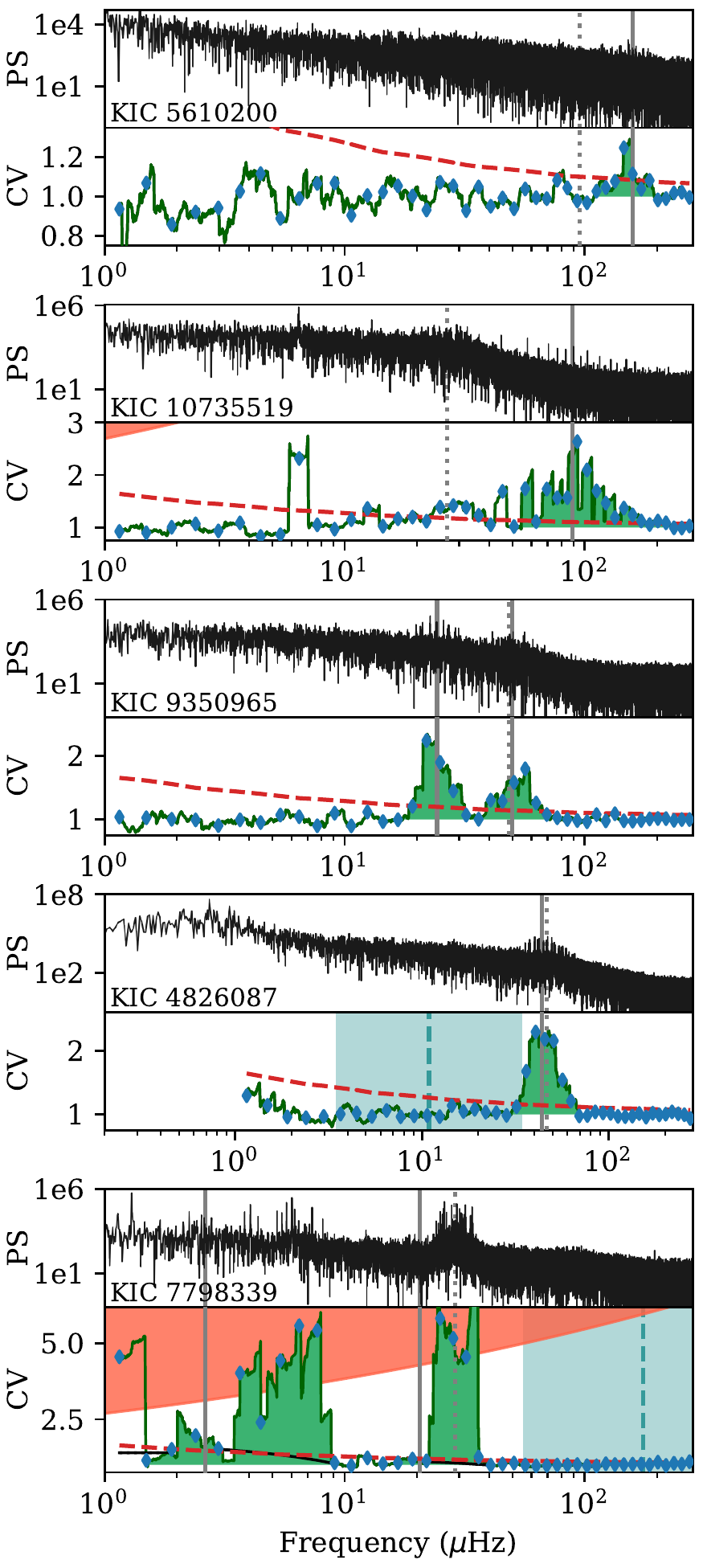}
	\caption{Power and CV spectra for targets, from top to bottom, KIC\,5610200, KIC\,10735519, KIC\,9350965, KIC\,4826087, and KIC\,7798339. The CV spectra are displayed as in Figures~\ref{fig:CVexample} and~\ref{fig:fp} (blue diamonds: independent CV; dark green curves: oversampled CV; black segments: interpolated CV; dashed red lines: 0.1\% FAP threshold; filled green; $\mathrm{CV} > 1$ regions around detections;  shaded red: upper CV limit; dotted vertical lines: \numax\ from APOKASC-2; solid vertical lines: measurements from our algorithm). The acceptable range of \numax\ given the estimate from FliPer (vertical dashed line) is shaded in blue for stars that are flagged for disagreement. We extend the displayed frequency range for KIC\,4826087 to reveal the power excess at a \numax\ below the range searched by our algorithm.  See text for discussion.}
	\label{fig:examples}
\end{figure}

The most conspicuous disagreement for a \numax\ value that carries no flags---marked with a square in the top panel of Figure~\ref{fig:results}---is KIC\,5610200. We display the power and CV spectra of this star as the top example in Figure~\ref{fig:examples}. The \numax\ value from the CV method (thick vertical line at 158.9\,$\mu$Hz) coincides with the oscillation signatures visible in the power spectrum, while the value from APOKASC-2 does not appear to match (dotted line at 95.4\,$\mu$Hz). The low mode heights observed suggest that the intrinsic \numax\ of this star may be above the long-cadence Nyquist frequency.

The most commonly returned flag is flag 1 (8.5\% of stars). This indicates that there are additional significant peaks in the independent CV spectrum besides any detected solar-like oscillations or spikes large enough to raise flag 2. Most \numax\ results that are accompanied by this flag still agree with the cataloged values to the same precision.  All results that are in disagreement with the APOKASC values have additional flags set (marked with crosses in the second panel of Figure~\ref{fig:results}), with the exception of KIC\,10735519 (marked with a square). We display the power and CV spectra of this target as the second example in Figure~\ref{fig:examples}. \citet{Zhou2010} characterized KIC\,10735519 as an Algol-type eclipsing binary with a 0.9070-day orbital period. The series of spikes caused by the binary signal increases the CV values above the expectation for $\chi^2_2$ noise through most of the spectrum, corrupting our automated \numax\ detection.

We displayed an example of a star with an anomalously high spike in its CV and power spectra in Figure~\ref{fig:CVexample} (KIC\,5793628; square in third panel of Figure~\ref{fig:results}). Situated near the pulsational power excess, this spike threatens to skew any \numax\ measurement if not accounted for. Figure~\ref{fig:CVexample} demonstrates how our algorithm avoids giving these spikes too much influence by interpolating across them in the oversampled CV spectra. Flag 2 is set for the 2.7\% of stars for which we perform this interpolation step. This correction explains why our \numax\ measurement for KIC\,5793628 is 26\% lower than the APOKASC value, which coincides with the spike frequency. All \numax\ measurements from spike-interpolated spectra that are more discrepant have other flags set.

Of the 69 targets (1.7\%) that our method returns multiple \numax\ candidates for (flag 3), 63 have two \numax\ values recorded, and six have three values.  KIC\,9350965, a star with two reported power excesses (two squares in the fourth panel of Figure~\ref{fig:results}),  is the third example in Figure~\ref{fig:examples}. Both detections correspond to solar-like oscillations in a blended light curve. The natural ability to find multiple power excesses is a strength of our approach, as it can enable the efficient search for asteroseismic binary systems, of which only three are known \citep[][]{Appourchaux2015,White2017,Beck2017}. Since APOKASC only lists one \numax\ for each target, additional \numax\ detections from our algorithm will necessarily disagree with the catalog, even when these correspond to real solar-like oscillation power excesses.

Flag 4 indicates that none of our returned \numax\ values are within 0.5\,dex of the estimate from FliPer, implying a disagreement between the power excess and the variance of the light curve expected from scaling relations. Only the 16 targets (0.4\%) in the fifth panel of Figure~\ref{fig:results} carry this flag. Our \numax\ values may not be good initial guesses for fitting the backgrounds of these power spectra. An interesting example is KIC\,4826087, whose variance is influenced by granulation corresponding to a pulsational power excess below our searched frequency range, near 0.7\,$\mu$Hz. We extend the displayed frequency range for this target in Figure~\ref{fig:examples}. There are few known examples of solar-like oscillations at such low frequencies \citep[e.g.,][]{Stello2014}.

Our last example in Figure~\ref{fig:examples} is the only star with the first four flags all set: KIC\,7798339 (marked with diamonds throughout Figure~\ref{fig:results}).  This is not a solar-like oscillator at all, but rather a $\gamma$ Doradus variable \citep{Uytterhoeven2011}.

The cataloged \numax\ values for the 11 stars that our algorithm does not return a detection for are marked in the bottom panel of Figure~\ref{fig:results}.  Five of these stars have $\nu_\mathrm{max}\leq 5\,\mu$Hz, where our significance threshold is highest and the mode lifetimes are not as well sampled. We also do not detect the Nyquist-aliased pulsations in another with an intrinsic $\nu_\mathrm{max}\approx 475\,\mu$Hz implied by APOKASC---its power excess is not obvious in the long-cadence observations. We are, however, able to recover the Nyquist-reflected \numax\ for a target with intrinsic $\nu_\mathrm{max}\approx 390\,\mu$Hz.

Our \numax\ measurements from CV spectra of the full sample of 6656 APOKASC-2 stars with KASOC data are tabulated in Appendix~\ref{app:tab}.  We also catalog the associated flags and emphasize that the vast majority of our \numax\ values are robust, even when accompanied by flags. We find that the 94 light curves that are flagged for having multiple candidate power excesses warrant visual inspection.  We note those that appear to the authors to genuinely display multiple solar-like oscillation power excesses in Appendix~\ref{app:multi}.

\section{Discussion and Conclusions}
\label{sec:fp}

Detecting the presence of solar-like oscillations is a prerequisite to red giant asteroseismology. The frequency of maximum pulsation power, \numax, is one of the fundamental seismic parameters of a solar-like oscillator. We have developed a rapid and robust method of detecting and measuring \numax\ of these oscillations in the presence of granulation backgrounds from coefficient of variation (CV) spectra. The \numax\ values obtained in our analysis of the APOKASC-2 giants typically agree with the cataloged values \citep{Pinsonneault2018} to an average precision of 2.7\% from four years of \Kep\ data. Our algorithm detects $>99\%$ of solar-like oscillations in giants with $<1\%$ false positives determined from a test on dwarf stars that do not exhibit visible pulsations in long-cadence data. Computing the CV spectra and searching for power excesses in 6656 APOKASC stars takes 70 seconds on a single CPU and is easily parallelized (most additional overhead is from reading in the time series and computing the power spectra). 

Our algorithm was designed to produce robust measurements in the presence of source blending, and it can even detect multiple power excesses in a single time series. Additional signals that may be scientifically interesting or obstructive can also be identified from CV spectra. For example, \citet{Colman2017} flagged \emph{by eye} 168 red giant power spectra that featured anomalous peaks, of which 110 overlapped with the APOKASC-2 sample that we analyzed. Our algorithm flags 103 of these as containing high spikes (66 targets) and/or additional significant CV values. Our method can quickly produce larger catalogs of anomalous features in solar-like oscillator power spectra, some of which \citet{Colman2017} argue are signatures of compact binaries orbiting the oscillating giants. The flags returned for all 6656 APOKASC-2 stars that we analysed are included with the \numax\ measurements in Appendix~\ref{app:tab}.

This data-driven approach to detecting \numax\ pairs naturally with methods such as the power spectrum of the power spectrum (e.g., \citealt{Hekker2010}; or equivalently, the autocorrelation of the time series; \citealt{Mosser2009}) to measure the large frequency separation, $\Delta\nu$.  This second asteroseismic parameter is usually easier to detect than \numax\ \citep[e.g.,][]{Chaplin2014}, and can be used as additional validation of a candidate power excess detection or to resolve whether a measured \numax\ is intrinsically super-Nyquist \citep{Chaplin2014b}. 

By providing a shortcut to \numax, the CV method will accelerate the process of fitting the granulation background to power spectra \citep[esp.\ where the detected \numax\ agrees with the prediction from FliPer;][]{Bugnet2017}. These fits could validate and refine our \numax\ values, and will benefit from knowledge of spike locations that the CV algorithm flags. This will shorten the time to bagging individual peaks for detailed asteroseismic analysis, and we plan to incorporate the CV method into the code of \citet{Andres2018}. The CV spectrum may also be a valuable data representation for training machine learning algorithms for detecting \numax, as has been attempted from the regular power spectrum by \citet{Hon2018}.

Ingredients of our approach have been applied to red giants and other astrophysical variables in the past.  \citet{Ransom2002} provide a detailed description of Fourier techniques that are relevant to the analysis of long time series.  In particular, \citet{Israel1996} took a similar approach to detecting coherent signals in coloured power spectra, motivated for the study of pulsars. The ``Moving-Windowed-Power-Search'' of \citet{Lund2012} also considers statistical excesses in the binned spectra of solar-like oscillators; however, it relies on first dividing by an optimized background model fit.  By combining elements of these approaches, we avoid the need to fit a background to reliably identify the presence and \numax\ of stochastic, solar-like oscillations.

We have analyzed here only a subset of the red giants observed by \Kep. \citet{Yu2018} recently identified 16{,}000 red giant stars in the \Kep\ data.  We plan to extend and optimize this algorithm for application to a larger \Kep\ sample, as well as to the CoRoT data and observations along the ecliptic from K2.  As currently defined, the algorithm encounters challenges in searching for power excesses in short time series; however, widening the bins used to calculate the CV spectra can lower the FAP significance thresholds at the cost of admitting more false positives.  We will explore these limitations and ways to mitigate them in future work, with the eventual goal of applying this detection technique to the upcoming data from TESS.

\section*{Acknowledgements}
The authors thank Nathalie Theme{\ss}l and Andres Garcia Saravia Ortiz de Montellano for valuable discussions about this work. Thanks go to Charlotte Kanngie{\ss}er for visually inspecting our data during her summer praktikum at MPS.  
This research is supported by the European Research Council under the European Community's Seventh Framework Programme (FP7/2007-2013) / ERC grant agreement no 338251 (StellarAges).
This paper includes data collected by the Kepler mission. Funding for the Kepler mission is provided by the NASA Science Mission directorate. 
The \emph{Kepler} light curves used in this work has been extracted using the pixel data following the methods described in \citet{Lund2015} and corrected following \citet{Handberg2014}.
This research has made use of the KASOC database, operated from the Stellar Astrophysics Centre (SAC) at Aarhus University, Denmark. 
Funding for the Stellar Astrophysics Centre (SAC) is provided by The
Danish National Research Foundation. 

\bibliographystyle{mnras}
\bibliography{bibliography}



\appendix

\section{Measurements of \numax\ for the APOKASC Sample of Giants}
\label{app:tab}

We applied the coefficient-of-variation algorithm for detecting solar-like oscillations (Section~\ref{sec:rgs}) to the KASOC-processed light curves available for 6656 stars from APOKASC-2 (see Section~\ref{sec:apokasc}). We provide the full output of our algorithm for every star in Table~\ref{tab:results} (full machine readable table available online). Stars with multiple candidate solar-like oscillation power excess detections are included once per \numax\ value. The binary bit flags returned by our algorithm (defined in Table~\ref{tab:flags}) are summed and presented as base-10 numbers in the ``flags'' column. The cataloged values from APOKASC-2 \citep{Pinsonneault2018} and the estimates from FliPer \citep{Bugnet2017} are included for comparison.

\begin{table}
	\centering
	\caption{CV-based \numax\ measurements for 6656 giants in APOKASC.  Full machine readable table available online.}
	\label{tab:results}
	\begin{tabular}{rcccr}
		\toprule
		KIC &      $\nu_\mathrm{max,meas}$ &    $\nu_\mathrm{max,cat}$ &  $\nu_\mathrm{max,FliPer}$ &  flags \\
		&  ($\mu$Hz)  &  ($\mu$Hz)  &  ($\mu$Hz)  &  \\
		\midrule
		8037095 &                 2.070 &     2.023 &       1.289 &  2 \\
		7672453 &                      &     2.023 &       1.910 &  0 \\
		7351928 &                 2.072 &     2.027 &       1.565 &  0 \\
		9697618 &                      &     2.028 &       1.290 &  3 \\
		2011145 &                 2.127 &     2.088 &       1.605 &  0 \\
		6615133 &                 2.247 &     2.119 &       2.043 &  0 \\
		8085217 &                 2.298 &     2.138 &       1.657 &  1 \\
		\bottomrule
	\end{tabular}
\end{table}

\section{Validation of multiple \numax\ detections}
\label{app:multi}

Out of 6656 light curves analysed, our algorithm returns multiple candidate \numax\ detections for 94. The authors inspected these light curves and their power spectra by eye to confirm which appear to exhibit two well-separated solar-like oscillation power excesses in reality. We list these 30 targets in Table~\ref{tab:multi}, with footnotes indicating those for which the second detection is not completely convincing or unique. Besides these stars with multiple sets of solar-like oscillations, some additional detections arise from, e.g., rotation signatures, remaining signal from photometric binaries, classical oscillations, and spurious detections from noise.

\begin{table}
	\centering
	\caption{APOKASC-2 targets that exhibit multiple solar-like oscillation power excesses.}
	\label{tab:multi}
	\begin{threeparttable}
		\begin{tabular}{rc}
			\toprule
			KIC &     notes   \\
			\midrule
			2161409 &  \\ 
			2422558 & a   \\ 
			2449518 &   \\ 
			2568888 & \\ 
			4173334 & \\ 
			4260884 & \\ 
			6206407 &  \\ 
			6501237 &  \\ 
			6689517 & a \\
			6888756 & \\ 
			7510604 &  \\ 
			8004637 &   \\ 
			8479383 &   \\ 
			8636389 &  \\ 
			9350965 &  \\ 
			9392650 &  \\ 
			9412408 &   \\ 
			9725292 &  \\ 
			9893437 &  b \\ 
			9893440 &  b \\ 
			10083224 &  \\ 
			10592924 &   \\ 
			10602015 &   \\ 
			10937954 &  \\ 
			10973854 &   \\ 
			11090673 & b   \\ 
			11090674 & b   \\ 
			11299484 &  \\ 
			12117790 & a \\
			12165692 &  \\ 
			
			\bottomrule
		\end{tabular}
		\begin{tablenotes}
			\item[a] Second power excess tentatively supported by visual inspection.
			\item[b] These same blends appear twice in the APOKASC sample.
		\end{tablenotes}
	\end{threeparttable}
\end{table}


\bsp	
\label{lastpage}
\end{document}